\documentstyle[multicol,aps,epsf]{revtex}

\begin{document}
\title{Mesoscopic effects in superconductor-ferromagnet-superconductor
 junctions}
\author{A.Yu. Zyuzin $^1$, B. Spivak $^2$, M. Hru\v{s}ka $^2$}
\address{ $^1$ A.F.Ioffe Physical-Technical institute,
194021 St.Petersburg,
Russia}
\address{ $^2$ Physics Department, University of Washington, Seattle, WA
98195}
\maketitle

\begin{abstract}
We show that at zero temperature the supercurrent through
the  superconductor - ferromagnetic metal -
superconductor junctions does not decay exponentially with the
 thickness $L$ of the junction.
At  large $L$ it has a random sample-specific sign which can change with
a change in temperature.
In the case of mesoscopic junctions the phase of the order parameter in the 
ground state is 
a random sample-specific quantity. In the case of junctions of large
area the ground state
phase difference is $\pm \pi/2$.
\end{abstract}

\pacs{Suggested PACS index category: }

\sloppy

This work has been motivated by recent experiments \cite{ryazanov,aprili}
on the superconductor-metallic ferromagnet-superconductor junctions where
the ground 
state of the system with the superconducting phase difference equal to
$\pi$ has been observed. It is well known that the sign of the critical
supercurrent of pure SFS junctions 
oscillates with the width $L$ of the ferromagnetic region 
\cite{buzdin,radovic,bergeret,nazarov}. This is due to the difference in
Fermi
momentum of electrons with different spins at the Fermi energy 
causing the superconducting wave function in the ferromagnetic
region to oscillate with the characteristic distance 
$\hbar/(k_{\uparrow}-k_{\downarrow})$.
Here $k_{\uparrow},k_{\downarrow}$ are Fermi momenta of up and down
spins, which are different because of the finite exchange spin-splitting
energy $I$ in ferromagnets.

It is important to note that in pure junctions the characteristic distance 
of oscillations is inversely
proportional to $I$ and that at zero temperature the modulus of the
critical current does not decay exponentially with $L$. 

In the opposite limit of disordered ferromagnets $L\geq
L_{I}=\sqrt{\frac{D}{I}}\geq l$
the average critical current $\langle I_{c}\rangle$
decays
exponentially
with 
$L$ and with temperature $T$  \cite{ryazanov}
 
\begin{equation}
\langle I_{c} \rangle \sim Re
\exp (-\frac{L}{\xi_{F}}); \xi_{F}=\sqrt{\frac{\hbar D}{2(\pi
T+iI)}}
\end{equation}
(where brackets $\langle \rangle$ stand for averaging over the random
realizations of the scattering potential and $D$ is the the electron
diffusion coefficient in the ferromagnet. 
It is interesting however, that in addition to the exponential decay, the
average critical current oscillates as a function of $L$. It also
changes sign as a
function of $T$, provided $D/L^{2} \stackrel{~}{<} I$. 

The value of $\langle I_{c}\rangle$ can be negative
which means that in the ground state of the junction the phase difference
$\phi$
of the superconducting order parameter is
$\pi$ rather than zero. The experimentally measured critical current in
this case is the absolute value $|I_{c}|$.  

The oscillations of the average critical current have been observed
experimentally \cite{ryazanov,aprili}.
We would like to note, however, that according to Eq.1 , these
oscillations
can be observed only in the case when the exchange spin-splitting energy
in the 
 ferromagnet is relatively small so that the characteristic distance of
oscillations
 is large.
This limits significantly the choice of ferromagnetic metals which can be
used in the junctions to observe this effect.
 
In this paper we show that the exponential decay of the supercurrent
 with $L$ in Eq.1 originates from the averaging procedure.
Before averaging over the impurity configurations 
at $L\gg L_{I}$ the supercurrent $I_{s}(\phi, L)$ has a {\it
random}
sample-specific sign while its  modulus does not decay
 exponentially.  
The practical consequence of this conclusion is that the Josephson 
effect survives in the case of ferromagnets with large $I$ when $L\gg
L_{I}$. We would like to mention that this feature is a
particular case of a general statement that the Friedel oscillations in
disordered metals do not decay exponentially at zero temperature
\cite{SpivakZyuzin,SpivakZyuzin1}.

Below we discuss the limit of thick SFS junctions $L\gg L_{I}$ when their
superconducting properties are determined by
mesoscopic effects. 
We show that in this case the critical current of the junctions {\it does
not depend} on
the spin splitting $I$. The ground state of such a junction is, generally
speaking, doubly degenerate with the phase difference 
having a random sample-specific modulus distributed 
in the interval ($0, \pi$).
We also show that the critical current undergoes random oscillations as 
a function of $T$. Thus the junctions with $L\gg L_{I}$ should
exhibit the same sequence of effects as the junctions with $L<L_{I}$.

The  energy of the Josephson junction $E(\phi)$ is an even and periodic
function of the phase of the order parameter. It can be represented
in the form
\begin{equation}
E(\phi)=\sum_{n=1}^{\infty}E_{n}\cos n\phi
\end{equation}
while the current through the contact is determined by the relation  
$J=2e\frac{d E}{d \phi}$.
The coefficients $E_{n}$ are random sample-specific quantities.
For $L\gg L_{I}$  all average coefficients
$\langle E_{n} \rangle \sim \exp (-L/L_{I})$ are exponentially small.
In this case the typical values of the coefficients $E_{n}$ can be
estimated from their variances. 
To simplify the analysis we consider the case when the superconductors
and the ferromagnet are separated by tunneling barriers of small
transparency.
We will show below that at $D/L^{2}\geq T$ and $L\geq \xi$ we have 
\begin{equation}
\langle (E_{1})^{2}\rangle =S\xi ^{2}(\frac{g}{8\pi \nu _{0}D})^{4}(%
\frac{D}{2\pi ^{2}L^{2}})^{2}
\end{equation}
\begin{equation}
\langle (E_{2})^{2} \rangle =S\xi^{2}(\frac{g}{8\pi
 \nu_{0}D})^{8} \frac{%
1}{(4\pi)^{3}}\frac{\xi^{2}D^{2}}{L^{2}}
\end{equation}
(where $S$ is the area of the junction, $\nu_{0}$ is the density of
states
per spin in the ferromagnet, $\xi$ is the zero-temperature coherence
length in the superconductor and $g$ is the conductance per unit area
of the surface between the ferromagnet and the superconductor. Eqs. 3,4
correspond to the diagrams shown in Figures 1.b),c).
In the case when the tunneling transmission coefficient of the insulator
between the
superconductor and the ferromagnet is small, the typical amplitude
$(\langle (E_2)^2 \rangle )^{1/2}$ of the
second harmonic contains an additional power of $g$ compared to the
amplitude $(\langle (E_1)^2 \rangle )^{1/2}$ of the first 
harmonic and therefore for typical samples of low $g$, $E(\phi)$ is
well approximated by the first harmonic. Since $(\langle (E_1)^2 \rangle
)^{1/2}$ does not decay exponentially with $L$, it is
much larger than 
$\langle E_{1} \rangle$.  This means that $E_{1}$ has a random sign.
In the cases when $E_{1}<0$ the ground state
 of the junction corresponds to $\phi_{GS}=\pi$.
 
The correlation function
$\langle E_{1} E_{2} \rangle$ contains an additional power of $g$ with
respect to $\langle ( E_{2}) ^{2}\rangle$ and can
be neglected which means that $E_{1}$ and $E_{2}$ are random uncorrelated
quantities.
 In rare samples where the amplitude of the first harmonic is small
 $(E_{1}\sim E_{2})$ 
one has to take into account the second harmonic in Eq.2.
 In this case
the ground state is doubly degenerate and 
corresponds to the phase difference 
\begin{equation}
\phi_{GS}\sim \pm \arccos \frac{E_{1}}{4 E_{2}}
\end{equation}
Therefore the sample-specific random absolute value of the ground state
phase
difference is in
the interval $0<|\phi _{GS}|<\pi/2$, provided $|E_{1}|<4|E_{2}|$.
The probability of such an event is of order
\begin{equation}
\sqrt{\frac{\langle ( E_{2}) ^{2}\rangle}{\langle ( E_{1}) ^{2}
\rangle}}
\sim \xi L(\frac{g}{8\pi \nu _{0}D})^{2}
\end{equation} 

 We would like to mention
 that in the
case of a transparent superconductor-ferromagnet boundary $\phi_{GS}$
is a random sample-specific quantity of order one. 

Let us now discuss the temperature dependence of the critical
 current $I_{c}(T)$.
To do so we calculate the correlation function
\begin{equation}
\langle E_{1}(T_{1}) E_{1}(T_{2}) \rangle \sim \exp (-2L\sqrt{%
\pi(T_{1}+T_{2})/D})
\end{equation}
It follows from Eq.7 that the quantities $\langle (E_{1}(T))^{2}\rangle
\sim
 \exp (-2^{3/2}L
\sqrt{T/D})$
and $\langle (E_{1}(T) E_{1}(0)\rangle \sim \exp (-2 L \sqrt{T/D})
\rangle$ decay with different rates as $T$ increases.
This indicates that in addition to an exponential decay, the quantity
$E_{1}(T)$ exhibits random sample-specific
 oscillations with a period
of order $T^{*}\sim D/L^{2}$.

The results presented above were obtained in the approximation when the
variations of the phase of the order parameter along the
superconductor-ferromagnet surface
are neglected. This is a good approximation for the samples of small area.
Below we show that in the samples of large area the
possibility of spatial fluctuations of the order parameter phase 
along the superconductor- ferromagnet surface leads to an
average critical current 
 which is proportional to the area of the
junction and does not decay exponentially even in the limit when $L\gg L_{I}$.
 In this case the ground state of the junction is doubly
degenerate and
\begin{equation}
\phi_{GS}=\pm \pi/2
\end{equation}
This can be shown using an expression for the effective Josephson energy
\begin{equation}
E=\int E_{c}({\bbox {\rho}}) \cos \phi ({\bbox {\rho}}) d {\bbox {\rho}}+
 \frac{N_{s}}{2 m}  \int d {\bf r} ({\bf \nabla}\phi({\bf r})^{2})
\end{equation}
where ${\bbox {\rho}}$ is the coordinate along the surface between the
superconductor and the ferromagnet, the ${\bf r}$-integration is
performed in the bulk of the superconductors, $m$ is the electron mass,
and  $N_{s}$ is the superfluid density in superconductors.
Eq.9 is valid on a scale larger than $L$ along the surface.
The random function $E_{c}({\bf y})$ is characterized by its average 
$\langle E_{c}({\bbox {\rho}}) \rangle=0$ and a quickly decaying (at
$|{\bbox {\rho}}-{\bbox {\rho '}}|>L$) correlation
function $\langle E_{c}({\bbox {\rho}}) E_{c}({\bbox {\rho '}})\rangle
\simeq $
$\langle E_{c}({\bbox {\rho - \rho '}}) E_{c}(0)\rangle \simeq \langle
(E_{1}(S=L^{2}))^{2}\rangle / L^4$. The
phase
difference
$\phi ({\bbox {\rho}})=
\langle \phi \rangle +\delta \phi ({\bbox {\rho}})$ is a random function
of
${\bbox {\rho}}$. At small $E_{1}$ $\delta \phi({\bbox {\rho}})\ll 1$.
Minimizing
Eq.9
with respect to $\delta \phi({\bbox {\rho}})$ we get an expression for the
effective energy per unit area \cite{pi/2}
 \begin{equation}
E_{eff}=-E_1 ^{eff}\sin^{2} \langle \phi \rangle
\end{equation}
(where $E_1^{eff} \simeq \frac{m}{N_s \xi L^2} 
\langle (E_1 (S=L^2))^2\rangle $),
determining the average critical current density as 
$\langle I_{c}\rangle = 2e E_1^{eff}$ and the average phase
difference in the ground
state as given by Eq.8.

The results in Eqs.3,4 were calculated microscopically describing the SFS
junction by a
Hamiltonian of the form 

\begin{equation}
\widehat{H}=\widehat{H}_{BCS}+\widehat{H}_{T}+\widehat{H}_{F}
\end{equation}
where $\widehat{H}_{BCS}$ is the Hamiltonian of superconducting leads,
the Hamiltonian 
\begin{equation}
\widehat{H}_{T}=t\sum\limits_{i=1,2;\alpha }\int_{S_{i}}
d{\bbox {\rho}}_{i}\left( \Psi _{i}^{+}\left( {\bbox
{\rho}}_{i},z_i;\alpha
\right) \Psi
_{F}\left( {\bbox {\rho}}_{i},z_i;\alpha \right) +h.c\right) 
\end{equation}
 describes tunneling between superconductors
labeled by $i=1,2$ and the  
ferromagnetic metal ($z_1 =0,z_2=L$), labeled by index $F$, and  $\alpha $
is the spin
index.
The integration is taken over the surfaces between superconductors
and the ferromagnetic metal. The last term in Eq.11 corresponds to
the disordered ferromagnetic
metal unperturbed by the presence of superconductors, where spin up and
down bands are split by the exchange field $I$ \ 

\begin{equation}
\widehat{H}_{F}=\widehat{H}_{0}+
I\int d{\bf r}\Psi _{F}^{+}\left( {\bf r};\alpha \right) {\bf \sigma }
_{\alpha \beta }^{z}\Psi _{F}\left( {\bf r};\beta \right) {\bf .}
\end{equation}

Here $\widehat{H}_{0}$ is the Hamiltonian of noninteracting electrons
which
contains the operators of the kinetic energy and a random field $U\left(
{\bf r}\right) $ . We assume that the random potential is white-noise
correlated so
$\left\langle U\left( {\bf r}%
\right) \right\rangle =0$ and $\left\langle U\left( {\bf r}\right) U\left( 
{\bf r}^{\prime }\right) \right\rangle =\left( 2\pi \nu _{0}\tau \right)
^{-1}\delta \left( {\bf r}-{\bf r}^{\prime }\right) $ (where  $%
\tau $ is the mean free time).

In the lowest order in tunneling through the superconductor-ferromagnet
boundary we get
\begin{equation}
E_{1}=\frac{T}{2}
\sum\limits_{\epsilon _{k};\alpha
}t^{4}\int_{S_{1};S_{2}}d{\bbox {\rho}}_{1}d{\bbox {\rho}}_{2}d{\bbox
{\rho}}_{1}^{\prime }d{\bbox {\rho}}_{2}^{\prime }\{F^{+}
\left( \epsilon _{k};{\bbox {\rho}}_{1},0;{\bbox {\rho}}_{1}^{\prime
},0\right)
G_{-\alpha ;-\alpha }\left( \epsilon _{k};{\bbox {\rho}}_{1}^{\prime
},0;{\bbox {\rho}}_{2}^{\prime },L\right) F \left( \epsilon _{k};{\bbox
{\rho}}_{2}^{\prime },L;{\bbox {\rho}}_{2},L\right) G_{\alpha ;\alpha }\left(
\epsilon
_{k};%
{\bbox {\rho}}_{2},L;{\bbox {\rho}}_{1},0\right) +h.c\}
\end{equation}
where $\epsilon_{k}=\pi (2k+1)T$ is the Matzubara frequency and
$k=0,1,...$ is an integer.
We use the usual diagrammatic technique for averaging the products of
electron Green's functions \cite{abrikosov}.
Diagrams for correlation functions $\left\langle \left(
E_{1}\right) ^{2}\right\rangle
$ and $\left\langle \left( E_{2}\right) ^{2}\right\rangle
$ are shown in Fig.1.b),c).
 There are two important blocks in these diagrams.
The first one corresponds to diffuson and cooperon ladder made of single
particle Green's functions in the
ferromagnet. In the case of a large $I$ only diffusons
 and cooperons for parallel spins survive.
They are equal to each other and the same as in the absence of the
external magnetic field, satisfying the equation \cite{altshuler}
\begin{equation}
\left( -D{\bf \nabla }^{2}+\left| \epsilon _{k}-\epsilon _{k^{\prime
}}\right| \right) C\left( \left| \epsilon _{k}-\epsilon _{k^{\prime
}}\right| ;{\bf r},{\bf r}^{\prime }\right) =\theta \left( -\epsilon
_{k}\epsilon _{k^{\prime }}\right) \delta \left( {\bf r}-{\bf r}^{\prime
}\right)
\end{equation}
We would also like to mention that for
$I\tau
\geq 1$ we can neglect contributions of diagrams shown in Fig.1.d). 
 
The second block is the ladder made of the 
anomalous Green's functions $F(\epsilon _{k};{\bf r},{\bf r}^{\prime })$
in the superconductor. This average can be expressed in terms of the
averaged product of the advanced and retarded Green's function in the
normal metal. Since
\begin{equation}
F(\epsilon _{k};{\bf r},{\bf r}^{\prime })=-i\int\limits_{-\infty }^{\infty }%
{\displaystyle{dE \over 2\pi }}%
{\displaystyle{\Delta  \over \Omega ^{2}+E^{2}}}%
\left( G^{R}\left( E,{\bf r},{\bf r}^{\prime }\right) -G^{A}\left( E,{\bf r},%
{\bf r}^{\prime }\right) \right)
\end{equation}
(where $\Omega ^{2}\equiv \Delta ^{2}+\epsilon _{k}^{2}$), 
the averaging $\left\langle F(\epsilon _{k};{\bf r},{\bf r}^{\prime
})F(\epsilon
_{k}^{\prime };{\bf r},{\bf r}^{\prime })\right\rangle $ is equivalent to
the averaging of Green's functions in the normal metal. The diffusion
propagator $\left\langle F(\epsilon _{k};{\bf r},{\bf r}^{\prime
})F(\epsilon _{k}^{\prime };{\bf r},{\bf r}^{\prime })\right\rangle \sim $ $%
P(\Omega ,\Omega ^{\prime };{\bf r},{\bf r}^{\prime })$ obeys
the diffusion equation

\begin{equation}
\left( -D{\bf \nabla }^{2}+\Omega +\Omega ^{\prime }\right) P(\Omega ,\Omega
^{\prime };{\bf r},{\bf r}^{\prime })=%
{\displaystyle{1 \over 2}}%
{\displaystyle{\Delta  \over \Omega }}%
{\displaystyle{\Delta  \over \Omega ^{\prime }}}%
\delta \left( {\bf r}-{\bf r}^{\prime }\right)
\end{equation}

 The result of integration of four electron Green's functions over the
surface
in diagrams shown in Fig.1.b),c)
is estimated as $\tau ^{2}g$ , where $g$ is the dimensional tunneling
conductance per unit area of the boundary.

The solution of Eq.15 satisfying the boundary conditions $%
{\displaystyle{d \over dz}}%
C\left( \left| \epsilon _{k}-\epsilon _{k^{\prime }}\right| ;{\bf r},{\bf r}%
^{\prime }\right) =0$ at the superconductor - ferromagnet metal surfaces 
is given by (
for $ z>z^{\prime }$ )

\begin{equation}
C\left( \left| \epsilon _{k}-\epsilon _{k^{\prime }}\right| ;{\bf r},{\bf r}%
^{\prime }\right) =\theta \left( -\epsilon _{k}\epsilon _{k^{\prime
}}\right) \int 
\frac{dq_x dq_y}{( 2\pi) ^{2}}
\frac{\cosh ( Q(z-L/2)) \cosh ( Q(z^{\prime }+L/2)) }
{DQ\sinh QL}
e^{i q_x (x-x') +i q_y (y-y')}
\end{equation}
where $Q=\sqrt{q^{2}+%
{\displaystyle{\left| \epsilon _{k}-\epsilon _{k^{\prime }}\right|  \over D}}%
}$ . Substituting this expression into diagrams shown in Fig.1.b)c) we
get
\begin{equation}
\left\langle \left( E_{1}\right) ^{2}\right\rangle \sim S\xi ^{2}(\frac{g}{%
8\pi \nu _{0}D})^{4}T^{2}\sum\limits_{\epsilon _{k};\epsilon _{k^{\prime
}}}\theta \left( -\epsilon _{k}\epsilon _{k^{\prime }}\right) \int 
\frac{d^{2}{\bf q}}{( 2\pi) ^{2}}
{\displaystyle{1 \over \left( Q\sinh QL\right) ^{2}}}%
\end{equation}

\begin{equation}
\left\langle \left( E_{2}\right) ^{2}\right\rangle \sim S\xi
^{4}D^{2}(\frac{%
g}{8\pi \nu _{0}D})^{8}T^{2}\sum\limits_{\epsilon _{k};\epsilon
_{k^{\prime
}}}\theta \left( -\epsilon _{k}\epsilon _{k^{\prime }}\right) \int
{\displaystyle{d^{2}{\bf q} \over \left( 2\pi \right) ^{2}}}%
{\displaystyle{1 \over \left( Q\sinh QL\right) ^{4}}}%
\end{equation}
Calculating integrals in Eqs.19,20 we arrive at Eqs.3-4.

To calculate the temperature dependence $\left\langle E_{1}\left(
T_{1}\right)
E_{1}\left( T_{2}\right) \right\rangle $ we should substitute $%
\epsilon _{k}=\pi \left( 2k+1\right) T_{1}$ and $\epsilon _{k^{\prime }}=\pi
\left( 2k+1\right) T_{2}$ into Eq.19 . Then for
$L_{T_{1}}=(D/T_{1})^{1/2},L_{T_{2}}=(D/T_{1})^{1/2}<L$ we get $%
\left( \sinh QL\right) ^{-2}= 4\exp [-2L(q^{2}+
(\pi T_{1}+\pi T_{2})/D)^{1/2}]$ and finally we arrive to Eq.7.

In conclusion we have shown that the critical current of a mesoscopic SFS
junction 
at small temperatures does not decay exponentially with the ferromagnet
thickness. It has a random sign, which changes with temperature. The
ground state
phase difference of the junction is a random quantity $0<\phi_{GS} <
\pi/2$.
In the case of junctions of large area the phase difference is
$\phi_{GS}=\pi/2$.
Let us estimate the typical value of the critical current using Eq.3.
Taking, for example, iron as a ferromagnet with $L\sim \xi=3\times 10^{-6} 
 $ cm, the
area
of the surface 
$S\sim 10^{-7}$ cm$^2$, the elastic mean free
 path $l\sim 10^{-6}$ cm and assuming that the transmission coefficient
through the superconductor-ferromagnet boundary is of order one we get 
$I_{c}\sim 10^{-6}$ A.
The estimate based on Eq.3 scales with the junction area as $\sqrt{S}$ and
is valid at small $S$. For junctions of large area, Eq.10 implies
that the critical current is
proportional to $S$. 

We express our thanks to M. Gershenzon, A.D.Kent and
Z. Radovi\'{c} for valuable discussions.

This work was supported by Division of Material Sciences, U.S.National
Science Foundation under Contract No. DMR-9205144 and (ZA) by Russian Fund
for Fundamental Research 01-02-17794.~

\newpage

\begin{figure}[!ht]
\centerline{\epsfxsize=12cm \epsfbox{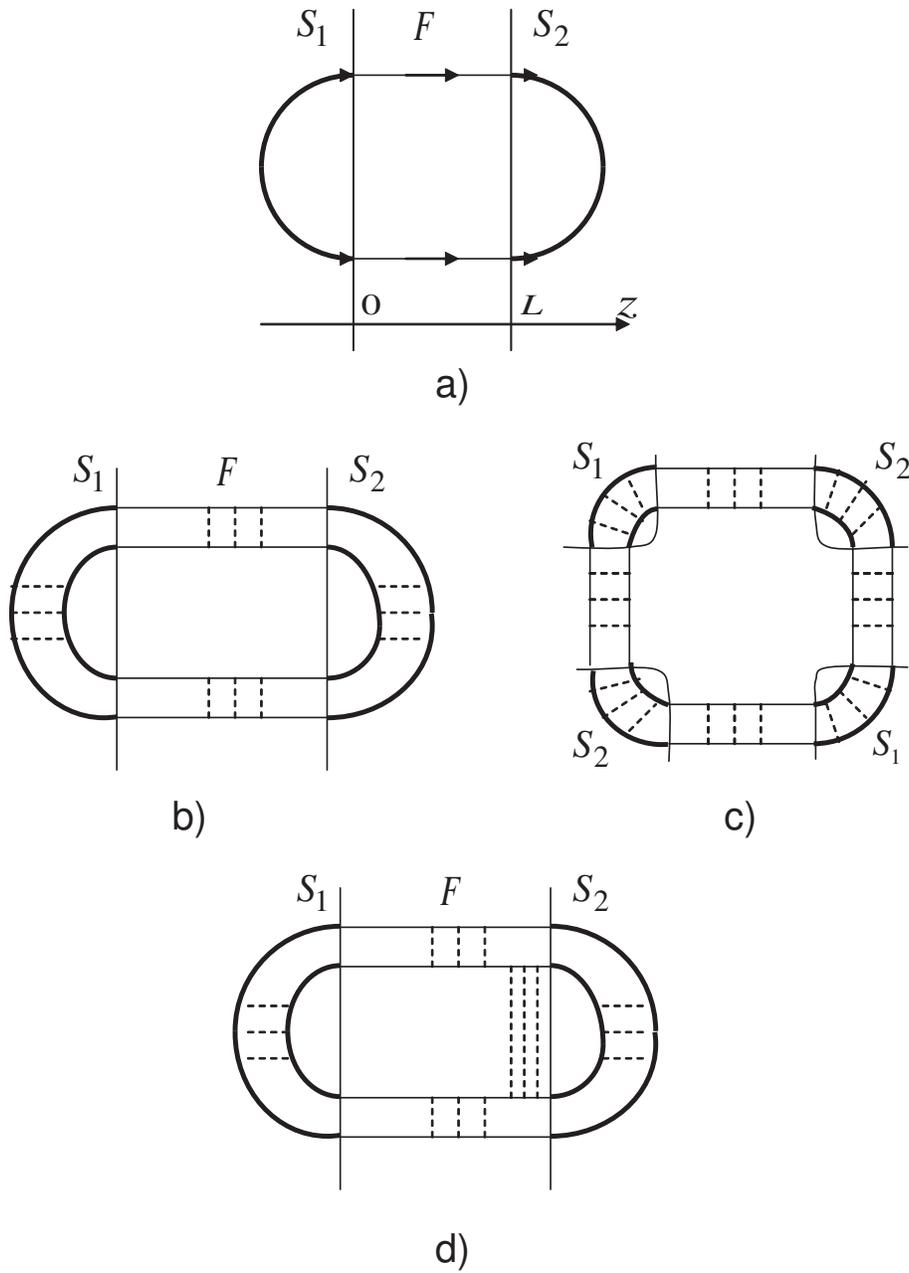}}
\caption{ The diagrams contributing to: a) the current, to
lowest order in transparency,  b) the correlator $\langle E_{1} E_{1}
\rangle,  $ c) the correlator
$\langle E_{2} E_{2} \rangle $,  d) a diagram insignificantly contributing
to the correlator $\langle E_{1} E_{1} \rangle $ in a superconductor -
ferromagnet - superconductor junction. The symbols $S_1$
and $S_2$ correspond to the first and second superconductor
respectively and $F$
denotes the ferromagnet. Thin lines
represent the single electron Green's functions, thick lines
 represent the anomalous Green's functions, dashed lines
denote averaging over the configurations of the impurity scattering
potential. Ladders formed of Green's functions and impurity scatterings in
the ferromagnet correspond to diffuson or cooperon ladders.} 
\end{figure}  

\end{document}